\documentclass{article}

\usepackage{amssymb,amsfonts,amsmath}
\usepackage{cite,enumerate,float,indentfirst}
\usepackage{color}

\def\be{\begin{eqnarray}}
\def\ee{\end{eqnarray}}
\def\nn{\nonumber}

\def\p{\partial}
\def\tr{{\rm tr}\,}
\def\Tr{{\rm Tr}\,}

\def\Hurwitz{Kronecker\ }
\def\Kroneker{Kronecker\ }

\definecolor{red}{rgb}{1,0,0}
\definecolor{orange}{rgb}{1,0.5,0}
\definecolor{violet}{rgb}{0.7,0,1}

\hoffset= - 1.0in         

\begin{document}

\title{\vspace{.1cm}{\LARGE {\bf From \Kroneker to tableau pseudo-characters in tensor models
}\vspace{.5cm}}
\author{{\bf H. Itoyama$^{a,b}$},
{\bf A. Mironov$^{c,d,e}$},
\ {\bf A. Morozov$^{d,e}$}
}
\date{ }
}

\maketitle

\vspace{-6.2cm}

\begin{center}
\hfill FIAN/TD-13/18\\
\hfill ITEP/TH-24/18\\
\hfill IITP/TH-14/18\\
\hfill OCU-PHYS-484
\end{center}

\vspace{4.cm}

\begin{center}
$^a$ {\small {\it Department of Mathematics and Physics, Graduate School of Science,
Osaka City University, 3-3-138, Sugimoto, Sumiyoshi-ku, Osaka, 558-8585, Japan}}\\
$^b$ {\small {\it Osaka City University Advanced Mathematical Institute (OCAMI), 3-3-138, Sugimoto, Sumiyoshi-ku, Osaka, 558-8585, Japan}}\\
$^c$ {\small {\it I.E.Tamm Theory Department, Lebedev Physics Institute, Leninsky prospect, 53, Moscow 119991, Russia}}\\
$^d$ {\small {\it ITEP, B. Cheremushkinskaya, 25, Moscow, 117259, Russia }}\\
$^e$ {\small {\it Institute for Information Transmission Problems,  Bolshoy Karetny per. 19, build.1, Moscow 127051 Russia}}\\
\end{center}

\vspace{.5cm}

\begin{abstract}
We present a brief summary of the recent discovery of direct tensorial analogue of characters.
We distinguish three degrees of generalization:
(1) $c$-number \Kroneker characters made with the help of
symmetric group characters and inheriting most of the
nice properties of conventional Schur functions,
except for forming a complete basis for the case of rank $r>2$ tensors: they are orthogonal, are eigenfunctions
of appropriate cut-and-join operators and
form a complete basis for the operators with non-zero Gaussian averages;
(2) genuine matrix-valued tensorial quantities, forming an
over-complete basis but difficult to deal with; and
(3) intermediate {\it tableau pseudo-characters},
depending on Young tables rather than on just Young diagrams,
in the \Kroneker case, and on entire representation matrices,
in the genuine one.
\end{abstract}

\bigskip

\bigskip

Characters play a prominent role in group theory and especially in its
physical applications,
because they take values in {\it numbers} and thus are more
comprehensible to physicists than generic objects in representation theory.
Still amusingly much can be expressed through them.
In particular, the simplest characters of the linear group $SL(\infty)$,
the Schur functions appear to be the best tool to describe the most important property of matrix models, their
{\it super}\,integrability,
the result of intersection of determinant representations (KP-integrability) \cite{KMMOZ}
and Virasoro-like constraints \cite{vircon}.
We refer to a recent \cite{MMreviewchar} for a detailed review of the subject
and references.

In this paper, we summarize a new important development \cite{IMMchar}
extending this story from matrix to tensor models,
which are attracting increasing attention for a variety of reasons \cite{tensor,tenfirst,GuraupostWit,KleTar,Gr,SY,K,R1,physlast,BGRfirst,Gur,Bonz,Virtree,GurVir,more,uncl,Ram,Cristo,
tenlast}.
Despite there are no more matrices and Lie algebras in this case,
very close counterparts of characters still exist,
and they play just the same role in describing  the {\it super}\,integrability \cite{IMM3}
of {\it rainbow} tensor models \cite{IMM1,IMM2,IMM3}, the ones with the highest possible
``gauge'' invariance.
Below is a very brief collection of theses, for details of motivations, formulas and examples see \cite{IMMchar}.
Our work has a serious overlap with \cite{Ramg2},
especially at the level of standard symmetric group technologies,
which are not yet well exposed in theoretical physics literature (see, however, \cite{MM2,Ramg1}).

\begin{itemize}

\item{Rectangular complex matrix model} (RCM) \cite{RCM}  can be defined
as an integral over complex matrices of the size $N_1\times N_2$
\be
Z_{N_1,N_2}^{\rm RCM}\{t\}
=\frac{1}{ {\rm Vol}_{U_N}} \int  \exp\Big(-\mu\, \Tr M\bar M + \sum_k t_k\,\Tr (M\bar M)^k\Big) \, dM
\ee
which is a Toda chain $\tau$-function satisfying the Virasoro constraints \cite{MMint,KMMOZ,UFN3}.
It is {\it super}integrable, i.e. all its Gaussian correlators can be calculated
{\it explicitly} \cite{IMM2,MM1}:
\be
\Big< \chi_R[M\bar M]\Big>=\frac{1}{ {\rm Vol}_{U_N}} \int dM e^{-\mu\Tr M\bar M} \chi_R[M\bar M]
=\mu^{-N_1N_2}\cdot \frac{\chi_R^*\{N_1\}\chi_R^*\{N_2\}}{d_R}
\label{gavSchur}
\ee
where $R$ is an arbitrary Young diagram, while
$\chi_R[M\bar M]=\chi_R\{p_k=\Tr (M\bar M)^k\}$  are the corresponding Schur functions,
i.e. the {\it characters} of the linear group $gl_N$
expressible through symmetric-group characters $\psi_R(\Delta)$
\be
\chi_R\{p\} = \sum_{\Delta \vdash |R|} \frac{\psi_R(\Delta)}{z_\Delta}\cdot p_\Delta
\label{Schurthroughpsi}
\ee
The sum at the r.h.s. runs over all Young diagrams $\Delta$
of the same size (number of boxes) $|R|$
as $R$, if $\Delta=[\delta_1\geq \delta_2\geq \ldots]$ has $m_i$ lines of length $i$ then
$p_\Delta := \prod_k p_{\delta_k} = \prod_i p_i^{m_i}$
and $z_\Delta := \prod_i i^{m_i}\cdot m_i!$ where the product runs over all $i$ with non-zero $m_i$.
The value of the Schur function at the special point, 
$\chi_R^*(N) := \chi_R\{p_k = N\}$ gives dimension of the representation $R$ of $gl_N$, and $d_R:=\chi_R\{\delta_{k,1}\}={1\over n!}\cdot{Dim}_R$ is related to dimension ${Dim}_R$ of the representation $R$ of the symmetric group $S_n$, $n=|R|$.

\item{Rainbow tensor model} is a direct generalization of RCM
with rectangular matrix substituted by a complex-valued tensor of rank $r$,
$M_{\vec a}=M_{a_1\ldots a_r}$ and $a_i=1\ldots N_i$.
It has ``gauge" symmetry $U(N_1)\otimes\ldots \otimes U(N_r)$.
For the sake of brevity, the $r$-tuples are often denoted by arrows.

\item{Gauge invariant operators in the rainbow  model}
are linear combinations of monomial operators, which are arbitrary convolutions
of $n$ tensors $M$ and $n$ tensors $\bar M$ (in this context $n$ is called "level")
and  are labeled by $r$ permutations $\sigma_i\in S_n$, elements of
the level dependent symmetric group $S_n$:
\be
{\cal K}^{\vec\sigma} :=  \prod_{p=1}^n M_{a_1^{(p)},\ldots,a_r^{(p)}}
\bar M^{a_{\sigma_1(p)}^{(p)},\ldots, a_{\sigma_p(p)}^{(p)} }
\ee
They are invariant under common left and right multiplications of all of the $r$
permutations,
\be
{\cal K}^{g_L\circ\vec\sigma\circ g_R} = {\cal K}^{\vec\sigma}
\ee
i.e.  are enumerated by the points of the double coset
$S_n\backslash S_n^r/S_n$, which has size
${\cal N}_r(n)=\sum_{\Delta\vdash n} z_\Delta^{r-2}$.

\item{The number of connected gauge invariant operators} ${\cal N}_r(n)$
actually coincides with the number of connected Feynman-'t Hooft diagrams ${\bf N}_{r-1}(n)$
with $n$ propagators in the ``previous rank" rainbow model:
\be
\boxed{
{\cal N}_r(n) = {\bf N}_{r-1}(n)
}
\label{FD}
\ee
In the particular case of {\it Aristotelian} model with $r=3$, this number ${\cal N}_3(n)$ is
the number of unlabeled Grothendieck's {\it dessins d'enfant} with $n$ edges \cite{Gro},
i.e. the number of Feynman-'t Hooft diagrams in the matrix RCM: $1,3,7,26,97,624,\, \ldots$

\item{We {\it define} the {\it \Hurwitz} tensorial character}, depending on the
$r$-tuple of Young diagrams as a direct generalization of
(\ref{Schurthroughpsi}):
\be
\boxed{
\chi_{\vec R}(M,\bar M) := \frac{1}{n!}\sum_{\vec\sigma\in S_n^{\otimes r}}
\psi_{\vec R}[\vec\sigma]\cdot {\cal K}^{\vec\sigma}
} \ = \frac{1}{n!}
\sum_{\sigma_1,\ldots,\sigma_r\in S_n} \psi_{R_1}[\sigma_1]\ldots
\psi_{R_r}[\sigma_r]\cdot {\cal K}^{\sigma_1,\ldots,\sigma_r}
\label{krochar}
\ee
where $n$ is the common size of all the Young diagrams $R_i$,
and $[\sigma]$ is the Young diagram, which describes
the cycle type (conjugacy class) of the permutation $\sigma$.

\item{The Gaussian average of this tensorial character} is easily calculated from
the result of \cite{MM2},
\be
\Big<{\cal K}^{\vec\sigma}\Big>  =
\!\!\!\!\!\sum_{Q_1,\ldots,Q_r\,\vdash\, n}\left( \sum_{\gamma\in S_n}\
\prod_{i=1}^r \psi_{Q_i}[\gamma\circ\sigma_i]\cdot\chi_{Q_i}^*(N_i)\right):= \sum_{\vec Q}\sum_{\gamma\in S_n}
\psi_{\vec Q}[\gamma\circ\vec\sigma] \cdot\chi_{\vec Q}^*(\vec N)
\label{aveK}
\ee
and is  given by a direct
generalization of (\ref{gavSchur}):
\be
\boxed{
\Big<\chi_{\vec R} \Big> = C_{\vec R}\cdot \prod_{i=1}^r {\chi_{R_i}^*(N_i)\over d_{R_i}}
}
\label{avechi}
\ee
where the Kroneker coefficients are
\be
C_{\vec R}
= \frac{1}{n!}\sum_{\gamma \in S_n} \prod_{i=1}^r \psi_{R_i}[\gamma]: = \frac{1}{n!}\sum_{\gamma \in S_n} \psi_{\vec R}[\gamma]
\label{Clebsh}
\ee
Note that these $C_{\vec R}= \sum_{\Delta\vdash n} \prod_{i=1}^r \psi_{R_i}[\Delta]$ are different from (dual to) the Hurwitz numbers
\be
{\bf N}_{\vec\Delta} = n!\ \sum_{R\vdash n} d_R^{2-2g-r} \prod_{i=1}^r {\psi_R(\Delta_i)\over z_{\Delta_i}} 
\ee
which count the number of ramified coverings and are related at $r=3$ to the number of
{\it dessins d'enfants} at the r.h.s. of (\ref{FD}).

\item{If $C_{\vec R}$ vanishes}, so does the character $\chi_{\vec R}$,
not only its Gaussian average:
\be
\boxed{
C_{\vec R} = 0 \ \Longrightarrow \ \chi_{\vec R} = 0
}
\label{vanavvanchi}
\ee
All non-vanishing characters $\chi_{\vec R}$ are linearly independent, and the set of $\chi$ is redundant in the space of Gaussian averages, i.e. there are non-vanishing linear combinations $\sum_{\vec R} a_{\vec R} \chi_{\vec R}\neq 0$
with vanishing averages, $\Big<\sum_{\vec R} a_{\vec R} \chi_{\vec R}\Big>= 0$. As soon as the characters $\chi_{\vec R}$ do not form a complete basis in the space of all gauge-invariant operators, there are non-vanishing gauge-invariant operators that are {\it not}
representable as sum of characters, but their Gaussian averages are:
${\cal O} \neq \sum_{\vec R} a_{\vec R} \chi_{\vec R}$, but
$\Big<{\cal O}\Big> = \Big<\sum_{\vec R} a_{\vec R} \chi_{\vec R}\Big>$.

\item{The \Hurwitz characters are common eigenfunctions of the generalized
cut-and-join operators} of \cite{MMN1}:
\be
\hat W^{\vec\sigma} \, \chi_{\vec R} = \lambda_{\vec R}^{\vec\sigma} \, \chi_{\vec R}
\ee
where
\vspace{-0.5cm}
\be
\hat W^{\vec\sigma} = {1\over n!}\ :{\cal K}^{\vec\sigma}\left(M,\frac{\p}{\p M}\right):
\ee
and the normal ordering implies that all the $M$-derivatives, which replace $\bar M$,
stand to the right of all $M$'s.
The eigenvalues $\lambda_{\vec R}^{\vec\sigma}$ are non-vanishing only when $C_{\vec R}\neq 0$ and, in this case,
\be
\lambda_{\vec R}^{\vec\sigma} =
\frac{\sum_{\gamma\in S_n}\psi_{\vec R}[\gamma\circ\vec\sigma]}
{\sum_{\gamma\in S_n}\psi_{\vec R}[\gamma]}
\ee

\item{\Hurwitz characters are orthogonal} in the sense that
\be
\hat \chi_{\vec R}\,\chi_{\vec R'} ={ \delta_{\vec R,\vec R'}\over d_{\vec R}}
\ee
where
\vspace{-0.5cm}
\be
\hat\chi_{\vec R} := \ :\chi_{\vec R}\left(M,\frac{\p}{\p M}\right):\
= \frac{1}{n!}\sum_{\vec\sigma} \psi_{\vec R}[\vec\sigma]\cdot\hat W^{\vec\sigma}
\ee
and $d_{\vec R}:=\prod_i d_{R_i}$.

\item{The \Hurwitz characters} $\chi_{\vec R}$ select combinations of gauge-invariant
operators with enhanced symmetry: while ${\cal K}^{\vec\sigma}$ are not invariant
under arbitrary conjugations of individual permutations $\sigma_i$,
the combinations $\chi_{\vec R}$ are, because the symmetric characters
$\psi_R[\sigma]$ depend only on the conjugacy classes of permutations.
Actually, at least at the lowest levels $n\leq 4$,
these $\chi_{\vec R}$ form a ring, i.e. the products of characters
\be
\chi_{\vec R}\cdot \chi_{\vec R'} = \sum_{R_i''\vdash |R_i|+|R_i'|}
A^{\vec R''}_{\vec R\vec R'}\,\chi_{\vec R''}
\ee
are expanded in characters just as it happens for the ordinary Schur functions.
Symmetric groups with $n>4$ are non-solvable
(have non-vanishing repeated commutators of arbitrary degree),
the Kroneker coefficients can exceed one,
and the number of gauge-invariant operators exceeds the number of Kroneker characters
(\ref{krochar}). The ring structure in this case is more sophisticated.

\item{As was already mentioned, the \Hurwitz characters} $\chi_{\vec R}$ form a redundant (overfull)
basis in the linear space of gauge-invariant operators with {\it non-vanishing}
Gaussian averages.

Perhaps, more important is that this basis is too small:
there are gauge-invariant operators ${\cal K}^{\vec \sigma}$
that can not be made from $\chi_{\vec R}$,
the size of the coset ${\cal N}_r(n)$ is much bigger than the
number of $r$-tuples of Young diagrams
(even if one neglects the additional constraint $C_{\vec R}\neq 0$).
The Gaussian averages of all these operators {\it are} linear combinations
of the averages of $\chi$, but operators themselves are {\it not}.
Thus, the \Hurwitz characters are not sufficient to form a full basis in the space
of {\it all} operators in the rainbow model with $r>2$.

\item{An overfull (even more redundant)
basis} in the space of all gauge-invariant operators can be made from the
more general quantities, which we call {\it Clebsh-Gordan}
\be
{\cal X}_{\vec R}(M,\bar M) :=
\frac{1}{n!}\sum_{\vec \sigma} {\cal S}_{\vec R}(\vec\sigma)\cdot {\cal K}^{\vec\sigma}
= \frac{1}{n!}
\sum_{ \sigma_1,\ldots,\sigma_r\in S_n}   \!\!\!
S_{R_1}(\sigma_1)\otimes\ldots\otimes S_{R_r}(\sigma_r)\cdot
{\cal K}^{\sigma_1,\ldots,\sigma_r}
\label{genchar}
\ee
where $S_R$ are matrices describing representations $R$ of the symmetric group $S_n$,
i.e. $S_R^{ij}(\sigma_1\circ\sigma_2)
= \sum_k S_R^{ik}(\sigma_1)S_R^{kj}(\sigma_2)$.
In variance with these tensor-valued quantities, the
\Hurwitz characters $\chi_R$ are
{\it numeric}:
they are made from traces,
\be
\psi_R[\sigma] = \tr S_R(\sigma)
\ee
which, additionally, depend only on the conjugacy classes $[\sigma]$ of $\sigma$,
and this explains why they are insufficient: operators ${\cal K}$ are {\it not}
invariants of most of these conjugations.

\item{
Since these matrices form a representation,
they can be chosen orthogonal for each particular $\sigma$ and $R$},
\be
\sum_{j=1}^{{\rm dim}_R} S_R^{ij}(\sigma)S_R^{kj}(\sigma) = \delta_{ij}
\ee
and they also satisfy more interesting orthogonality relations,
which hold for any finite group:
\be\label{orth}
\sum_{\sigma}S_R^{ij}(\sigma)S_{Q}^{kl}(\sigma)=\frac{\delta_{RQ}\delta^{ik}\delta^{jl}}{d_R}
\ \ \ \Longrightarrow \ \ \
\frac{1}{n!} \sum_{\sigma} \psi_R[\sigma]\psi_{Q}[\sigma] =
\sum_{\Delta\vdash n} \frac{\psi_R[\Delta]\psi_{Q}[\Delta]}{z_\Delta} =
\delta_{RR'}
\ee
because dimension ${\rm dim}_R$
of representation $R$ of $S_n$ is $d_R\cdot n!$

It follows from (\ref{orth}) that
\be
\sum_{\sigma \in S_n} S_R^{ij}(\sigma)S_Q^{kl}(\gamma\circ\sigma) =
\sum_m S_Q^{km}(\gamma) \sum_{\sigma \in S_n} S_R^{ij}(\sigma)S_Q^{ml}(\sigma)
= {1\over d_R}S^{ik}(\gamma^{-1})\cdot\delta^{jl}\delta_{RQ},
\ee
and, taking the trace of $S_Q$, we get
\be
\sum_{\sigma \in S_n} S_R^{ij}(\sigma)\cdot \psi_Q[\gamma\circ\sigma]
= {1\over d_R}S^{ij}_R(\gamma^{-1})\cdot\delta_{RQ}
\label{orthSpsi}
\ee

\item{As a manifestation of the above mentioned completeness of the \Hurwitz
{\it Gaussian averages}
and as  a direct corollary of (\ref{orthSpsi}), the Gaussian averages} of  ${\cal X}$
are immediately reduced to those of the \Hurwitz characters:
\be
\boxed{
\Big<{\cal X}_{\vec R}^{\vec i\vec j} \Big> =
\sum_{\vec\sigma} {\cal S}_{\vec R}(\vec\sigma)\cdot\Big<{\cal K}^{\vec\sigma}\Big>
\ \stackrel{(\ref{aveK})}{=}\
\sum_{\vec\sigma} {\cal S}_{\vec R}^{\vec i\vec j}(\vec\sigma) \sum_{\vec Q}
\sum_{\gamma\in S_n} \psi_{\vec Q}(\gamma\circ\vec\sigma)\cdot
\chi_{\vec Q}^*(\vec N) =
{\cal C}_{\vec R}^{\vec i\vec j}\cdot {\chi^*_{\vec R}(\vec N)\over d_{\vec R}}
}
\ee
The only difference from (\ref{avechi}) is that the Kroneker coefficients
get substituted by more sophisticated tensor-valued quantities
\be
{\cal C}^{i_1j_1|\ldots|i_rj_r}_{R_1\ldots R_r} =
{\cal C}^{\vec i\vec j}_{\vec R}
= \frac{1}{n!} \sum_{\gamma\in S_n} S^{\vec i\vec j}_{\vec R}(\gamma)
= \frac{1}{n!} \sum_{\gamma\in S_n} S^{i_1j_1}_{R_1}(\gamma)
\ldots S^{i_rj_r}_{R_r}(\gamma)
\ee
In particular, for $r=2,3$,
\be
{\cal C}^{i_1j_1|i_2j_2}_{R_1R_2}={1\over n!}\sum_\gamma S_{R_1}^{i_1j_1}(\gamma)S_{R_2}^{i_2j_2}(\gamma)
\stackrel{(\ref{orth})}{=}{1\over \hbox{dim}_{R_1}}\delta_{R_1R_2}\delta_{i_1i_2}\delta_{j_1j_2}\nonumber\\
{\cal C}^{i_1j_1|i_2j_2|i_3j_3}_{R_1R_2R_3}={1\over n!}\sum_\gamma S_{R_1}^{i_1j_1}(\gamma)S_{R_2}^{i_2j_2}(\gamma)
S_{R_3}^{i_3j_3}(\gamma)=\sum_\alpha\left(
\begin{array}{ccc}
R_1&R_2&R_3\\
i_1&i_2&i_3
\end{array}
\right)_\alpha\cdot\left(
\begin{array}{ccc}
R_1&R_2&R_3\\
j_1&j_2&j_3
\end{array}
\right)_\alpha
\ee
where $\left(
\begin{array}{ccc}
R_1&R_2&R_3\\
j_1&j_2&j_3
\end{array}
\right)_\alpha$ are $3j$-symbols (Clebsh-Gordan coefficients), and the subscript $\alpha$ is in charge of equivalent representations emerging when $C_{R_1R_2R_3}$ is non-unit.
This explains the name {\it Clebsh-Gordan} for the quantities (\ref{genchar}).
In fact, they are very close to {\it generic} $\tau$-functions of \cite{gentau},
but this goes too far beyond the frame of the present text.

As functions of $\vec R$, the structure constants ${\cal C}$
are proportional to the Kroneker coefficients
$C_{\vec R}$ in (\ref{Clebsh}),
\be
\boxed{
{\cal C}^{\vec i\vec j}_{\vec R} \sim C_{\vec R}
}
\ee
in accordance with the $C\neq 0$ selection rule for non-vanishing Gaussian averages.
However, the property (\ref{vanavvanchi}) is {\it no} longer correct:
vanishing ${\cal C}$ does {\it not} imply vanishing ${\cal X}$,
and, as was already mentioned, in variance with the \Hurwitz characters,
the genuinely tensorial ones
(\ref{genchar}) are expected to
form a full (and redundant) basis
in the space of {\it all} gauge-invariant operators.

\item{Since ${\cal K}^{\vec\sigma}$ are invariant under the common
left and right multiplications of {\it all} $\sigma$,
we can deduce from  (\ref{orth}) that}
\be
{\cal X}_{R_1R_2}^{ij|kl}(M,\bar M)
=\sum_{\sigma_1,\sigma_2}S_{R_1}^{ij}(\sigma_1)S_{R_2}^{kl}(\sigma_2)
\cdot{\cal K}^{\sigma_1,\sigma_2}
=\sum_{\sigma_1,\sigma_2}S_{R_1}^{ij}(\sigma_1)S_{R_2}^{kl}(\sigma_1^{-1}
\circ\sigma_2)\cdot {\cal K}^{1,\sigma_2}=
\nonumber\\
=\sum_{\sigma_1,\sigma_2}S_{R_1}^{ij}(\sigma_1)S_{R_2}^{nk}(\sigma_1)S_{R_2}^{nl}(\sigma_2)
\cdot {\cal K}^{1,\sigma_2}=
{\delta_{jk}\delta_{R_1R_2}\over d_{R_1}}\sum_{\sigma}S_{R_2}^{il}(\sigma)
\cdot{\cal K}^{id,\sigma}
\ee
and similarly for higher $r$, e.g. at $r=3$,
\be
{\cal X}_{R_1R_2R_3}^{i_1j_1|i_2j_2|i_3j_3}(M,\bar M)
=\left(\sum_{\sigma}S_{R_1}^{i_1j_1}(\sigma)S_{R_2}^{n_2i_2}(\sigma)S_{R_3}^{n_3i_3}(\sigma)\right)
\cdot\sum_{\sigma_1,\sigma_2}S_{R_2}^{n_2j_2}(\sigma_2)S_{R_3}^{n_3j_3}(\sigma_3)\cdot
{\cal K}^{id,\sigma_2,\sigma_3}
\ee
Clearly, if we leave indices $i$ and $j$ free (do not make convolutions),
the coefficients are {\it not} invariant under individual conjugations
of $\sigma_2$ and $\sigma_3$, what was an extra hidden symmetry
of the \Hurwitz characters.
But actually they are also not invariants of the {\it common} conjugation,
which {\it is} the symmetry of ${\cal K}^{id,\vec\sigma}$.
This makes the set of tensorial characters ${\cal X}$ unnecessarily large.

\item{
In order to decrease redundancy, one can consider an intermediate object between the concise},
but insufficient, \Hurwitz characters
$\chi_{\vec R}$, and the huge matrix-valued ${\cal X}_{\vec R}$, that is, the objects
made from  the Young tableau $T$,
which we  call tableau  pseudo-characters:
\be
\boxed{
{X}_{\vec T}(M,\bar M) :=
\sum_{\vec \sigma} \Psi_{\vec T}(\vec\sigma)\cdot {\cal K}^{\vec\sigma} =
\!\!\!\!\!\sum_{\sigma_1,\ldots,\sigma_r\in S_n}\!\!\!
\Psi_{T_1}(\sigma_1) \ldots  \Psi_{T_r}(\sigma_r)\cdot
{\cal K}^{\sigma_1\,\ldots\,\sigma_r}
}
\label{tabchar}
\ee
They are number-valued, as $\chi$, but depend on additional data
(the tableau $T$ instead of the diagram $R=[T]$ {\it per se})
and are no longer invariant under individual conjugations of $\sigma_i$,
i.e. are capable to distinguish between ${\cal K}^{\{\gamma_i\sigma_i\gamma_i^{-1}\}}$
with different $i$-dependent $\gamma_i$ (while the \Hurwitz characters $\chi$ are obtained
by irreversible averaging over all $\gamma_i$'s).

\item{To define $\Psi_{T}(\sigma)$ and clarify the definition,
we remind a small piece from representation theory.}
The fact that the collection of matrices $S(\sigma)$
forms a representation of the group $S_n$ can be nicely
expressed in terms of its group algebra elements
\be
\hat{\cal S}_R^{ij} := \sum_{\sigma\in S_n} S_R^{ij}(\sigma)\cdot \sigma
\ee
\vspace{-0.5cm}
that satisfy
\be
\hat{\cal S}_R^{ij} \circ \hat{\cal S}_{R'}^{kl} = \sum_{\sigma,\sigma'} S_R^{ij}(\sigma)S_{R'}^{kl}(\sigma')
\cdot (\sigma\circ\sigma') =
\sum_{\sigma,\sigma'} S^{ij}_R(\sigma) S^{kl}_{R'}(\sigma^{-1}\circ\sigma')\cdot\sigma'
= \nn \\
=  {\sum_{\sigma,\sigma'} S^{ij}_R(\sigma)S^{km}_{R'}(\sigma^{-1})}S^{ml}_{R'}(\sigma')\cdot\sigma'
= {\sum_\sigma S_R^{ij}(\sigma)S_{R'}^{mk}(\sigma)}
\,\cdot\, \hat{\cal S}_{R'}^{ml} =
\frac{\delta_{RR'}\delta^{jk}}{d_R}\cdot \hat{\cal S}_R^{il}
\label{gramult}
\ee
Hence, one suffices to construct these elements of the group algebra $\hat{\cal S}_R^{ij}$ instead $S^{ij}_R(\sigma)$. In order to do this \cite{SG}, one can start from constructing primitive idempotents $\hat\Psi_T$
which are in one-to-one correspondence with Young tableau $T$,
and are called Young symmetrizers.
For a given Young tableau $T$, the Young symmetrizer is constructed as a product
\be
\hat\Psi_{T_i}:=\prod_{a}(\circ)\,\hat {\cal A}_a \circ \prod_b(\circ)\hat \Sigma_b=\sum_\sigma\Psi_{T_i}(\sigma)\cdot\sigma
\ee
of operations $\Sigma_a$ of symmetrization of all elements
in the $a$-th line of $T_i$ and of operations $A_b$ of antisymmetrization
of all elements in the $b$-th row of $T_i$, considered as elements of the group algebra.
The so constructed element of the group algebra $\Psi_{T_i}$
is a primitive idempotent, the corresponding left module generates an irreducible representation $R=[T_i]$ associated with the form of $T$. In fact, since the number of the Young tableau $T_i$ of a given form $R$ coincides with the dimension of $R$, one has construct the elements $\hat{\cal S}_R^{ij}$ in terms of $T_i$. To this end, one has to associate with each Young tableau $T_i$ the set of sub-tableau elements of the group algebra $\Psi_{T_i}^{(a)}$ such that $T_i^{(n)}=T_i$, $T_i^{(n-1)}$ is obtained by removing the point $n$, $T_i^{(n-2)}$ is obtained by removing the points $n$, $n-1$ etc. Introduce also $d_{T_i^{(a)}}$ associated with the form of the Young Tableau $T_i^{(a)}$ and $\xi_{T_i}^{(a)}$ given recursively
\be
\xi_{T_i}^{(n-1)}=d_{T_i^{(n-1)}}\ \xi_{T_i^{(n-2)}}\circ\Psi_{T_i^{(n-1)}}\circ\xi_{T_i^{(n-2)}}\\
\xi_{T_i}^{(n-2)}=d_{T_i^{(n-2)}}\ \xi_{T_i^{(n-3)}}\circ\Psi_{T_i^{(n-2)}}\circ\xi_{T_i^{(n-3)}}\\
\ldots\\
\xi_{\Box}=Id
\ee
Now $\hat{\cal S}_R^{ij}$ satisfying (\ref{gramult}) can be manifestly constructed as
\be\label{snb}
\hat{\cal S}_R^{ij}=\xi_{T_i}^{(n-1)}\circ\Psi_{T_i}\circ\sigma_{ij}\circ\xi_{T_j}^{(n-1)}
\ee
where $\sigma_{ij}$ denotes the permutation that maps the Young tableau $T_i$ to $T_j$.

In fact, formula (\ref{snb}) describes the seminormal, not orthogonal basis of representation matrices. However, one can further construct a basis of orthogonal matrices, \cite{SG}.

\end{itemize}

\bigskip

\bigskip

\noindent
{\bf To conclude,} this paper fulfils one of the promises of \cite{MMreviewchar}:
that the property
\be
\boxed{\Big<{\rm character}\Big> = character
}
\label{chacha}
\ee
deeply analyzed there,
is not specific for matrix models, but survives  non-trivial generalizations.
This paper concerns a generalization from matrix to tensor models,
i.e. to generic {\it non-linear algebra} in the sense of \cite{NLA}, far beyond traditional group and representation theory.
This paper provides a short summary of relevant ideas, definitions and facts,
of which the most important are put in boxes.
We explained that (\ref{chacha}) is literally true in tensor models,
if the Gaussian average is taken of the \Hurwitz characters (\ref{krochar}).
Moreover, they form a full basis in the space of {\it characters}
$\chi_{\vec R}^*(\vec N)$ at the r.h.s.
However, the \Hurwitz characters are not quite enough to describe all gauge-invariant
operators, and as a step towards such description we suggested to consider 
{\it tableau} pseudo-characters (\ref{tabchar}).
Another interesting approach could be to consider non-Gaussian averages,
i.e. ignore the grading w.r.t. the level $n$.
For more details, see \cite{IMMchar}.

\section*{Acknowledgements}

Our work is partly supported by the grant of the Foundation for the Advancement of Theoretical Physics ``BASIS" (A.Mor. and A.Mir.), by  RFBR grants 16-01-00291 (A.Mir.) and 16-02-01021 (A.Mor.), by joint grants 17-51-50051-YaF, 18-51-05015-Arm (A.M.'s). The work of H.I. was supported by JSPS KAKENHI Grant Number JP15K05059. Support from JSPS Bilateral Joint Projects (JSPS-RFBR collaboration) ``Topological Field Theories and String Theory: from Topological Recursion
to Quantum Toroidal Algebra'' from MEXT, Japan is appreciated.


\begin{thebibliography}{12}

\bibitem{KMMOZ} S. Kharchev, A. Marshakov, A. Mironov, A. Orlov, A. Zabrodin,
Nucl.Phys., {\bf B366} (1991) 569-601

\bibitem{vircon} F. David, Mod.Phys.Lett. {\bf A5} (1990) 1019\\
A. Mironov, A. Morozov, Phys.Lett. {\bf B252} (1990) 47-52\\
J. Ambj{\o}rn, Yu. Makeenko, Mod.Phys.Lett. {\bf A5} (1990) 1753\\
H. Itoyama, Y. Matsuo, Phys.Lett. {\bf 255B} (1991) 20

\bibitem{MMreviewchar} A. Mironov and A. Morozov,
 arXiv:1807.02409

\bibitem{IMMchar}
H. Itoyama, A. Mironov and A. Morozov, {\it to appear}

\bibitem{tensor} F. David, Nucl.Phys. {\bf B257} (1985) 45\\
V.A. Kazakov, I.K. Kostov, A.A. Migdal,
Phys.Lett. {\bf B157} (1985) 295\\
J. Ambjorn, B. Durhuus, T. Jonsson, Mod.Phys.Lett. {\bf A6} (1991) 1133-1146\\
N. Sasakura, Mod.Phys.Lett. {\bf A6} (1991) 2613\\
P. Ginsparg, hepth/9112013\\
M. Gross, Nucl.Phys.Proc.Suppl. {\bf 25A} (1992) 144-149

\bibitem{tenfirst} E. Witten, arXiv:1610.09758

\bibitem{GuraupostWit} R. Gurau, Nucl. Phys. {\bf B916} (2017) 386, arXiv:1611.04032; arXiv:1702.04228

\bibitem{KleTar} I. Klebanov, G. Tarnopolsky, Phys.Rev. {\bf D 95} (2017) 046004, arXiv:1611.08915\\
S. Carrozza, A. Tanasa, Letters in Mathematical Physics, {\bf 106(11)} (2016) 1531-1559, 1512.06718

\bibitem{Gr} D. Gross, V. Rosenhaus, JHEP, {\bf 02} (2017) 093, arXiv:1610.01569; {\it ibid.}, {\bf 05} (2017) 092, arXiv:1702.08016; arXiv:1706.07015; arXiv:1710.08113\\
Ch. Krishnan, S. Sanyal, P.N. Bala Subramanian, JHEP {\bf 03}
(2017) 056, arXiv:1612.06330\\
F. Ferrari, 	arXiv:1701.01171\\
V. Bonzom, L. Lionni, A. Tanasa, J.Math.Phys. {\bf 58} (2017) 052301, arXiv:1702.06944\\
M. Beccaria, A.A. Tseytlin, arXiv:1703.04460

\bibitem{SY}  S. Sachdev, Y. Ye, 
Phys.Rev.Lett. {\bf 70} (1993) 3339, cond-mat/9212030\\
J. Polchinski, V. Rosenhaus,
JHEP, {\bf 04} (2016) 001, arXiv:1601.06768\\
W. Fu, D. Gaiotto, J. Maldacena, S. Sachdev, Phys.Rev. {\bf D95} (2017)
026009,
arXiv:1610.08917\\
M. Berkooz, P.Narayan, M. Rozali, J. Simon, 
arXiv:1610.02422

\bibitem{K} A. Kitaev, "A simple model of quantum holography",
http://online.kitp.ucsb.edu/online/entangled15/kitaev/, http:
//online.kitp.ucsb.edu/online/entangled15/kitaev2/. Talks at KITP, April
7, 2015 and May 27, 2015\\
S. Sachdev, 
Phys.Rev. {\bf X5} (2015) 041025, arXiv:1506.05111\\
A. Jevicki, K. Suzuki, J. Yoon, JHEP, {\bf 07} (2016) 007,
arXiv:1603.06246\\
J. Maldacena and D. Stanford, 
arXiv:1604.07818\\
D. Bagrets, A. Altland, A. Kamenev, 
Nucl.Phys. {\bf B911} (2016) 191-205, arXiv:1607.00694\\
A. Jevicki, K. Suzuki, JHEP, {\bf 11} (2016) 046,
arXiv:1608.07567

\bibitem{R1} Z. Bi, C.-M. Jian, Y.-Z. You, K.A. Pawlak, C. Xu, arXiv:1701.07081\\
S.-K. Jian, H. Yao, arXiv:1703.02051\\
S. Carrozza, V .Lahoche, D. Oriti, arXiv:1703.06729\\
Ch. Krishnan, K. Pavan Kumar, S. Sanyal, JHEP, {\bf 06} (2017)
036, arXiv:1703.08155\\
M. Casali, P. Cristofori, S. Dartois, L. Grasselli, arXiv:1704.02800\\
Ch. Peng, JHEP, {\bf 05} (2017) 129, arXiv:1704.04223\\
S. Das, A. Jevicki, K. Suzuki, arXiv:1704.07208\\
Ch. Krishnan, K.V. Pavan Kumar,	
arXiv:1706.05364\\
Ch. Krishnan, K.V. Pavan Kumar, D. Rosa, arXiv:1709.06498; arXiv:1804.10103\\
S. Das, A. Ghosh, A. Jevicki, K. Suzuki, JHEP {\bf 1807} (2018) 184, arXiv:1712.02725; JHEP {\bf 1802} (2018) 162, arXiv:1711.09839\\
X.-H. Ge, S.-J. Sin, Yu Tian, Sh.-F. Wu, Sh.-Y. Wu, JHEP 1801 (2018) 068, arXiv:1712.00705\\
S. Okumura, K. Yoshida, Nucl.Phys. B933 (2018) 234-247, arXiv:1801.10537\\
Ch. Peng, arXiv:1805.09325 

\bibitem{physlast}
H. Kyono, S. Okumura, K. Yoshida, arXiv:1704.07410\\
J. Yoon,  	
arXiv:1706.05914,
arXiv:1707.01740\\
Ch. Peng, M. Spradlin, A. Volovich,
arXiv:1706.06078\\
T. Azeyanagi, F. Ferrari, F. Schaposhnik Massolo,
arXiv:1707.03431\\
K. Bulycheva, I. Klebanov, A. Milekhin, G. Tarnopolsky,
arXiv:1707.09347

\bibitem{BGRfirst}
R. Gurau, Commun.Math.Phys. {\bf 304} (2011) 69-93, arXiv:0907.2582; Annales Henri Poincare {\bf 11} (2010) 565-584, arXiv:0911.1945;
Class.Quant.Grav. {\bf 27} (2010) 235023, arXiv:1006.0714; Annales Henri Poincare {\bf 13} (2012) 399–423, 1102.5759\\
J.B. Geloun, R. Gurau, V. Rivasseau, Europhys.Lett. {\bf 92} (2010) 60008, arXiv:1008.0354

\bibitem{Gur} R. Gurau, V. Rivasseau, Europhys.Lett. {\bf 95} (2011) 50004, arXiv:1101.4182\\
R. Gurau, J.P. Ryan, SIGMA {\bf 8} (2012) 020, arXiv:1109.4812

\bibitem{Bonz} V. Bonzom, R. Gurau, V. Rivasseau, Phys.Rev. {\bf D85} (2012) 084037, arXiv:1202.3637\\
R. Gurau; V. Rivasseau; S. Gielen, L. Sindoni; J.P. Ryan; V. Bonzom; S. Carrozza; T. Krajewski, R. Toriumi; A. Tanasa; SIGMA {\bf 12} (2016): ''Special Issue on Tensor Models, Formalism and Applications", http://www.emis.de/journals/SIGMA/Tensor\_Models.html

\bibitem{Virtree} V. Bonzom, R. Gurau, A. Riello, V. Rivasseau, 
Nucl.Phys. {\bf B853} (2011) 174-195, arXiv:1105.3122\\
R. Gurau, 
Nucl.Phys. {\bf B852} (2011) 592, arXiv:1105.6072

\bibitem{GurVir} R. Gurau, arXiv:1203.4965\\
V. Bonzom, arXiv:1208.6216

\bibitem{more} V. Bonzom, JHEP {\bf 06} (2013) 062, arXiv:1211.1657\\
V. Bonzom, F. Combes, arXiv:1304.4152\\
V. Bonzom, R. Gurau, J.P. Ryan, A. Tanasa, JHEP, {\bf 09} (2014) 05, arXiv:1404.7517\\
R. Gurau, A. Tanasa, D.R. Youmans, Europhys.Lett. {\bf 111} (2015) 21002, arXiv:1505.00586

\bibitem{uncl} A. Tanasa, J.Phys. A: Math.Theor. {\bf 45} (2012) 165401, 1109.0694; SIGMA {\bf 12} (2016) 056, 1512.02087\\
S. Dartois, V. Rivasseau, A. Tanasa, Annales Henri Poincare, {\bf 15} (2014) 965-984, 1301.1535

\bibitem{Ram} D. Garner, S. Ramgoolam, Nucl.Phys. {\bf B875} (2013) 244-313, arXiv:1303.3246

\bibitem{Cristo} P. Cristofori, E. Fominykh, M. Mulazzani, V. Tarkaev, arXiv:1609.02357

\bibitem{tenlast}
Materials of the 2nd French-Russian Conference on Random Geometry and Physics
(2016), http://www.th.u-psud.fr/RGP16/

\bibitem{IMM1} H. Itoyama, A. Mironov, A. Morozov, Phys.Lett. B771 (2017) 180-188, arXiv:1703.04983

\bibitem{IMM2}
H. Itoyama, A. Mironov, A. Morozov, JHEP, 06 (2017) 115, arXiv:1704.08648 

\bibitem{IMM3}
H. Itoyama, A. Mironov and A. Morozov, Nucl.Phys. B932 (2018) 52-118, arXiv:1710.10027

\bibitem{Ramg2} J. Ben Geloun, S. Ramgoolam, arXiv:1708.03524

\bibitem{MM2} A. Mironov and A. Morozov, Phys.Lett. B774 (2017) 210-216, arXiv:1706.03667

\bibitem{Ramg1}
R. de Mello Koch, S. Ramgoolam, arXiv:1002.1634 \\
J. Ben Geloun, S. Ramgoolam, arXiv:1307.6490;  arXiv:1806.01085\\
P. Diaz, S.J. Rey, arXiv:1706.02667,  arXiv:1801.10506\\
R. de Mello Koch, D. Gossman, L. Tribelhorn, JHEP, 2017 (2017) 011, arXiv:1707.01455\\
P. Diaz,  arXiv:1803.04471

\bibitem{MM1}
A. Mironov, A. Morozov, Phys.Lett. B771 (2017) 503-507, arXiv:1705.00976

\bibitem{RCM}
T. Morris, 
Nucl.Phys. {\bf b356} (1991) 703-728\\
Yu. Makeenko, Pis'ma v ZhETF {\bf 52} (1990) 885-888\\
Yu. Makeenko, A. Marshakov, A. Mironov, A. Morozov, Nucl.Phys. {\bf B356} (1991) 574-628

\bibitem{MMint} A. Gerasimov, A. Marshakov, A. Mironov, A. Morozov, A. Orlov,
Nucl.Phys. {\bf B357} (1991) 565-618\\
S. Kharchev, A. Marshakov, A. Mironov, A. Morozov,
Nucl.Phys. {\bf B397} (1993) 339-378, hep-th/9203043

\bibitem{UFN3}
A. Morozov, Phys.Usp.(UFN) 37 (1994) 1, hep-th/9303139; hep-th/9502091; hep-th/0502010\\
A. Mironov, Int.J.Mod.Phys. {\bf A9} (1994) 4355, hep-th/9312212; Phys.Part.Nucl.
{\bf 33} (2002) 537; hep-th/9409190

\bibitem{Gro}
G. Belyi, Mathematics of the USSR: Izvestiya, {\bf 14:2} (1980) 247-256\\
A. Grothendieck, {\sl Sketch of a Programme}, Lond. Math. Soc. Lect.
Note Ser. {\bf 242} (1997) 243-283;
{\sl Esquisse d'un Programme}, in: P. Lochak, L. Schneps (eds.), Geometric Galois Action, pp.5-48,
Cambridge University Press, Cambridge (1997)\\
G.B. Shabat, V.A. Voevodsky,
The Grothendieck Festschrift, Birkhauser,
1990, V.III., p.199-227\\
N. Adrianov, N. Amburg, V. Dremov, Yu. Levitskaya, E. Kreines, Yu. Kochetkov, V. Nasretdinova, G. Shabat, arXiv:0710.2658

\bibitem{MMN1} A. Mironov, A. Morozov, S. Natanzon,  Theor.Math.Phys. {\bf 166} (2011) 1-22,
arXiv:0904.4227;  Journal of Geometry and Physics {\bf 62} (2012) 148-155,
arXiv:1012.0433

\bibitem{gentau} A. Mironov, A. Morozov, L. Vinet,
Teor.Mat.Fiz. \textbf{100} (1994) 119-131 (Theor.Math.Phys. {\bf 100} (1995) 890-899),
hep-th/9312213;\\
A. Gerasimov, S. Khoroshkin, D. Lebedev, A. Mironov, A. Morozov,
Int.J.Mod.Phys. \textbf{A10} (1995) 2589-2614,
hep-th/9405011;\\
S. Kharchev, A. Mironov, A. Morozov,
q-alg/9501013;\\
A.Mironov, hep-th/9409190; Theor.Math.Phys. {\bf 114} (1998) 127, q-alg/9711006

\bibitem{SG} D.E. Littlewood, {\sl The theory of group characters and
matrix representations of groups}, Oxford, 1958\\
M. Hammermesh, {\sl Group Theory and Its Application to Physical Problems (Dover Books on Physics)}, Dover Publications, 1989\\
G.D. James, A. Kerber, {\sl The Representation Theory of the Symmetric Group (Encyclopedia of Mathematics and its Applications)}, Cambridge University Press, 2009

\bibitem{NLA}
 I. Gelfand, M. Kapranov, A. Zelevinsky,
{\sl Discriminants,
 Resultants and Multidimensional Determinants},
Birkhauser, 1994 \\
V. Dolotin, A. Morozov,
{\sl Introduction to Non-Linear Algebra}, WS, Singapore 2007,  hep-th/0609022

\end{thebibliography}
\end{document}